\documentclass{elsart}

\usepackage{amssymb,amsmath,amsfonts,graphicx}

\begin{document}

\begin{frontmatter}



\title{Complexity analysis of the stock market}


\author[kaist]{Joongwoo Brian
Park,\thanksref{cor0}}
\author[kaist]{Jeong Won Lee,\thanksref{cor0}}
\author[yang]{Jae-Suk Yang,}
\author[jo]{Hang-Hyun Jo,\corauthref{cor1}}\ead{h2jo@kias.re.kr}
\author[kaist]{Hie-Tae Moon}
\corauth[cor1]{Corresponding author.}
\thanks[cor0]{The first two authors contributed equally to this work.}
\address[kaist]{Department of Physics, Korea Advanced Institute of Science and Technology,
Daejeon 305-701, Republic of Korea}
\address[yang]{Department of Physics, Korea University,
Seoul 136-713, Republic of Korea}
\address[jo]{School of Physics, Korea
Institute for Advanced Study, Seoul 130-722, Republic of Korea}
\begin{abstract}

We study the complexity of the stock market by constructing
$\epsilon$-machines of Standard and Poor's 500 index from February
1983 to April 2006 and by measuring the statistical complexities. It
is found that both the statistical complexity and the number of
causal states of constructed $\epsilon$-machines have decreased for
last twenty years and that the average memory length needed to
predict the future optimally has become shorter. These results
support that the information is delivered to the economic agents and
applied to the market prices more rapidly in year 2006 than in year
1983.

\end{abstract}

\begin{keyword}
econophysics \sep computational mechanics \sep statistical
complexity
%\PACS 87.19.Xx \sep 89.75.Hc \sep 87.23.Ge \sep 05.65.+b
\PACS 89.65.-s \sep 89.65.Gh \sep 89.75.-k

\end{keyword}

\end{frontmatter}

% main text

\section{Introduction}

Financial systems have been one of active research fields for
physicists. This interdisciplinary research area called econophysics
has been investigated by means of various statistical methods, such
as the correlation function, multifractality, minimal spanning tree,
minority games, continuous-time random walks, and spin models
\cite{Arthur1997,Mantegna2000,Bouchaud2000,Mandelbrot2001,Kaizoji2006,Kullmann2000,Challet1997,Scalas2000,Giada2002}.
Recently many empirical time series in financial markets become
available and has been also investigated by the rescaled range (R/S)
analysis to test the presence of correlations \cite{Peters1991} and
detrended fluctuation analysis to detect long-range correlations
embedded in seemingly non-stationary time series
\cite{Peng1994,Liu1999} and so on.

In this paper we adopt the computational mechanics (CM)
\cite{Feldman1998,Shalizi2001} to investigate the complexity of the
stock market. The CM is based on the early works of the information
and computation theory done by Shannon, Kolmogorov, and Chaitin
\cite{Shannon1948,Kolmogorov1965,Chaitin1966}. Despite its strong
functionality, CM has been applied only to analyze the abstract
models such as cellular automata \cite{Hanson1997,Shalizi2004} and
Ising spin system \cite{Crutchfield1997}, or empirical data in the
geomagnetism \cite{Clarke2003} and in the atmosphere
\cite{Palmer2000}. We believe that CM enables the complexities and
structures of different sets of data to be quantifiably compared and
that it directly discovers intrinsic causal structure within the
data \cite{Clarke2003}. This approach also shows how to infer a
model of the hidden process that generated the observed behavior.

We examined the tick data of Standard and Poor's 500 (S\&P500) index
from February 1983 to April 2006 by constructing deterministic
finite automata called ``epsilon-machine'' \cite{Crutchfield1989}
from the financial time series and by calculating the statistical
complexity from the constructed machine. The $\epsilon$-machine
captures the patterns and regularities in the observations in a way
that reflects the causal structure of the process. With this model
in hand, we can extrapolate beyond the original observations to
predict future behavior \cite{Shalizi2001}. The constructed
$\epsilon$-machine is a step toward the eventual use of such machine
in finding effective patterns embedded in the price index of stock
market. This is a novel approach to predict the next action on the
stock market with statistical probabilities. We also analyzed the
result that the complexity of the stock market has decreased.

\section{Principles}

According to Feldman \cite{Feldman1998} and Shailizi
\cite{Shalizi2001} we introduce the basics regarding to the
$\epsilon$-machine and the statistical complexity as complexity
measure.

\subsection{$\epsilon$-machine}

We consider a stochastic process given by an infinitely consecutive
discrete random variables, $\overleftrightarrow{X}=\cdots
X_{-1}X_0X_1X_2\cdots$, where each $X_i$ may take a symbol $x_i$
drawn from a finite countable set $A$ of size $k$. At any time $t$
this sequence of random variables can be divided into two
semi-infinite halves; a history $\overleftarrow{X_t}$ and a future
$\overrightarrow{X_t}$. If the process is conditionally stationary,
\textit{i.e.} for all possible future events $F$,
$\Pr(\overrightarrow{X_t}\in F\mid
\overleftarrow{X_t}=\overleftarrow{x})$ does not depend on $t$, then
we drop the subscript. And $\overrightarrow{X^l}$ and
$\overleftarrow{X^l}$ denote the first $l$ variables of
$\overrightarrow{X}$ and the last $l$ variables of
$\overleftarrow{X}$, respectively.

A causal state is defined as a set of history events (shortly,
histories) that have the same distribution of conditional
probabilities for all possible future events. $\epsilon$ is a
function that maps from histories to sets of histories:
\begin{equation}
\epsilon(\overleftarrow{x})=\{\overleftarrow{x'}\mid
\Pr(\overrightarrow{X}=\overrightarrow{x} \mid \overleftarrow{X}
=\overleftarrow{x})= \Pr(\overrightarrow{X} =\overrightarrow{x} \mid
\overleftarrow{X} =\overleftarrow{x'}), \forall \overrightarrow{x}
\in F\}.\label{eq:states}
\end{equation}
Each causal state $s_i$ consists of its name $i$, a set of histories
$\epsilon(\overleftarrow{x})$, and the conditional probability
distribution $\Pr(\overrightarrow{X}\in F\mid
\overleftarrow{X}=\overleftarrow{x})$, which is called ``morph.''
$S$ denotes the corresponding random variable and $\bold{S}$ does
the set of all causal states. Then, the transition probability
$T_{ij}^{(a)}$ is defined as the probability of generating a symbol
$a\in A$ when making the transition from state $s_i$ to state $s_j$;
\begin{equation}
T_{ij}^{(a)}\equiv \Pr(\overleftarrow{X}a\in s_j\mid
\overleftarrow{X}\in s_i),
\end{equation}
where $\overleftarrow{X}a$ is read as a semi-infinite sequence
obtained by concatenating $a\in A$ onto the end of
$\overleftarrow{X}$. Equivalently,
\begin{equation}
T_{ij}^{(a)}=\Pr(S' =s_j,\overrightarrow{X^1}=a \mid S=s_i),
\end{equation}
where $S$ and $S'$ are random variables for the current causal state
and its successor, respectively. The combination of the function
$\epsilon$ mapping from histories to causal states with the labeled
transition probabilities $T_{ij}^{(a)}$ is called the
$\epsilon$-machine, which represents a computational model
underlying the given time series. The causal states and the
transitions of $\epsilon$-machine form a directed graph, therefore
there can exist some states being never returned once the system
left those states. These are called transient states that cannot be
the true causal states so removed and the others are recurrent
states \cite{Shalizi2004b}. Once the $\epsilon$-machine is
constructed and the current causal state is identified, one can
optimally predict the future behavior of the process with some
conditional probability distributions, which will be useful in
practice, for example, for traders in financial markets.

For the operational applications the length of histories to be
considered is limited as $L_{\textrm{max}}$ for a process with
finitely consecutive random variables or finite time series.
$L_{\textrm{max}}$ should be large enough to fully detect the
structure embedded in the process. On the other hand it is also
limited by the total number $N$ of data points available in a way of
$L_{\textrm{max}}< \log_k N$ \cite{Shalizi2004b} for the significant
analysis.

\subsection{Statistical and topological complexities}

From the constructed $\epsilon$-machine, for each $i$ the
probability of finding the system in the $i$-th causal state after
the machine has been running infinitely long, $\Pr(s_i)$, can be
calculated. The components $T_{ij}$ of the transition matrix
$T=\sum_{a\in A} T_{ij}^{(a)}$ gives the probability of a transition
from state $s_i$ to state $s_j$. $\Pr(s_i)$ are obtained by solving
the following:
\begin{equation}
\sum_{i} \Pr(s_i)T_{ij}=\Pr(s_j).
\end{equation}
Then the statistical and topological complexities are defined as
\begin{equation}
C_\mu \equiv -\sum_{i} \Pr(s_i) \log_2 \Pr(s_i), \label{eq:Cmu}
\end{equation}
\begin{equation}
C_0 \equiv \log_2 \|\bold{S}\|,\label{eq:C0}
\end{equation}
where $\|\cdot\|$ represents the cardinality of a set. $C_\mu$
measures the minimum amount of historical information required to
make optimal forecasts \cite{Crutchfield1997,Palmer2000}. By the
definitions of the statistical and topological complexities, the
topological complexity is the upper bound of the statistical
complexity. And the equality holds when the distribution is uniform,
that is, for all causal states $\Pr(s_i)=1/\|\bold{S}\|$. As the
probability distribution of causal states deviate from uniformity,
the statistical complexity becomes smaller and therefore far from
the topological complexity.

\subsection{Simple examples}

Before closing this section, a few simplest examples are examined by
constructing the $\epsilon$-machines. The first is a process
generating the same symbol infinitely, such as $\cdots 000 \cdots$.
There exists the only one causal state consisting of the only one
history $\overleftarrow{x}=\cdots 000$ and the morph
$\Pr(0\mid\overleftarrow{x})=1$. If we depict each causal state as a
node and each transition from state $i$ to $j$ generating a symbol
$a$ with probability $p$ as an arc starting from a node $i$ to $j$
on which `$a\mid p$' is labeled, then the above $\epsilon$-machine
would be depicted as one node and one arc going back to itself with
`$0\mid 1$' labeled. In this case both complexities of Eqs.
(\ref{eq:Cmu}-\ref{eq:C0}) become $0$ bit.

The second example is a periodic series with period $2$, such as
$\cdots 0101 \cdots$. Two histories are found:
$\overleftarrow{x_1}=\cdots 101$, $\overleftarrow{x_2}=\cdots 010$.
Since $\Pr(0\mid\overleftarrow{x_1})=1$ while
$\Pr(0\mid\overleftarrow{x_2})=0$, each histories constitutes each
causal state. The constructed $\epsilon$-machine consists of two
nodes and two arcs coming from one circle to the other,
respectively. In this case both complexities of Eqs.
(\ref{eq:Cmu}-\ref{eq:C0}) become $1$ bit.

As the final one we toss a fair coin and record $H$ for head, $T$
for tail, and get a random process such as $\cdots HHTHT \cdots$.
There are infinitely many histories but all the morphs are the same
as
$\Pr(H\mid\overleftarrow{x_i})=\Pr(T\mid\overleftarrow{x_i})=1/2$.
Therefore the only one causal state is enough for the
$\epsilon$-machine, which consists of one node and two arcs going
back to itself but with different labels, `$H\mid 1/2$' and `$T\mid
1/2$', respectively. Both complexities of Eqs.
(\ref{eq:Cmu}-\ref{eq:C0}) become $0$ bit, which means the totally
random process is not complex at all. Thus, these measures satisfy
the so-called `boundary condition' for the complexity measure that
vanishes in the extreme ordered and disordered limits
\cite{Feldman1998b}.

\section{Empirical data analysis}

For the tick-by-tick S\&P500 index data from February 1983 to April
2006, as shown in Fig. 1, the statistical and topological
complexities are calculated from constructed $\epsilon$-machines
\cite{CSSR}. By using a time window of one year and shifting the
window by one month, we get $267$ data sets and for each data set
each $\epsilon$-machine is constructed. For convenience each data
set is named after its starting month, for example, the data set for
one year since February 1983 is called February 1983 data. The
average number of data points in one minute varies from about $1$ in
the early 1980's to $4$ in recent years. For the analysis we firstly
set a countable set $A$ to the smallest set of size $k=2$, such as
$\{0,1\}$. Then the original index data $Y_n$ change into the binary
time series $F_n$ by the following process:
\begin{equation}
F_n \equiv \theta \left(Y_{n+1}-Y_n\right),\label{eq:Fn}
\end{equation}
where $\theta(x)$ is a Heaviside step function. $F_n$ gets the value
of $0$ if the next index has decreased and does the value of $1$
otherwise. Since New York Stock Exchange opens during the day time,
we use only the intra-day data to avoid the discontinuous jump
between the previous day's closing index and the next day's opening
index due to overnight effects. In other words we exclude $F_n$ for
the difference between the last index of the previous day and the
first index of the next day.

To construct $\epsilon$-machines we set $L_{\textrm{max}}$ to $6$,
which gives the most reliable results. Figures 2 and 3 depict the
$\epsilon$-machines for the February 1983 data and for the April
2005 data, respectively. It is noteworthy that the number of causal
states has decreased for last twenty years, which will be discussed
later. In those figures, as mentioned, each numbered node represents
a causal state, while each arc joining one node to another does the
transition from one causal state to another. Each arc is labeled
with `$a\mid p$', that is, a symbol $a$ is generated with
probability $p$ by that transition. For example, in Fig. 3 if the
current state of the system is the $0$th one, the system goes back
to the $0$th state by generating symbol $1$ with probability
$p=0.634742$, while the system makes a transition to the $1$st state
by generating symbol $0$ with probability $1-p$. The directions of
these arcs tell us which causal state will be followed.

In more details, we investigate the histories belonging to each
causal state. The histories of each causal state for two
$\epsilon$-machines mentioned before are shown in Tables 1 and 2,
respectively. Since we set the length of the longest histories to be
considered as $6$, there are $2^6=64$ possible histories of length
$6$. In particular the histories of the $0$th causal state in Fig. 3
can be found in the first row of Table 2. If we add symbol $1$ on
the right end of each history and limit the length to $6$ (e.g.,
$000011 \rightarrow 000111$), we can see that all the resulting
histories remain included in the $0$th causal state. For the
opposite case of adding symbol $0$ on the right end (e.g., $000011
\rightarrow 000110$), all the resulting histories are found in the
$1$st causal state. By making use of this method repeatedly as we
want, we can predict the next finitely consecutive symbols with a
certain probability, which can be obtained just by multiplying the
transition probabilities along the arcs.

Next, we found that both the statistical and topological
complexities of S\&P500 index have a tendency of decreasing through
time as shown in Fig. 4. Since the time window is set to one year,
it is assumed that many short term events in the stock market, such
as the Black Monday, do not affect our analysis. Therefore we focus
on the long term behaviors of both complexities. Since the
difference between the statistical and topological complexities is
not significant for the whole range of times, the probability
distributions for causal states are almost uniform throughout time.
Conclusively our main concern is reduced to the decrease in the
number of causal states. Precisely, the total number of causal
states decreases from $42$ for the February 1983 data to $4$ for the
April 2005 data.

To find the underlying principle of the decrease in the number of
causal states through time, we revisit Tables 1 and 2 showing the
histories of length $L=6$ of the causal states for the February 1983
and April 2005 data, respectively. In Table 1 the $62$ histories are
mapped to $42$ causal states according to their morphs (the other
$2$ histories were in the removed transient states) and thus the
causal states are composed of only one to three histories. It is
found that for each causal state with two histories each two
histories are the same except the left end symbols of them, that is,
$L=5$ is enough to identify such states. But for the majority of
causal states $L=6$ is necessary. Therefore it is reasonable to
conclude that the average memory length we need to predict the
future at the February 1983 data was $6$, which corresponds to about
$6$ minutes.

In Table 2, the causal states for the April 2005 data are more
simplified than the above case. $64$ histories are grouped into $4$
causal states exactly containing $16$ histories, respectively. In
each causal state all the histories have the last two symbols in
common; $11$ for the $0$th causal state and $10$ for the $1$st one
and so on. The first $4$ symbols of each history are all the
possible matches of $0$'s and $1$'s. In this case the average memory
length to predict the future for the April 2005 data was $2$,
\textit{i.e.} about one half minute. In conclusion, the average
memory length needed to predict the future has decreased from $6$ in
the early 1980's to $2$ in recent years.

Now the decreasing tendency of statistical and topological
complexities for last twenty years is explained. We call the common
part of histories in each causal state an `effective pattern.' If
the length of effective pattern $L_{\textrm{eff}}$ decreases, the
number of possible effective patterns decreases as
$2^{L_{\textrm{eff}}}$, so does that of the resultant causal states.
Although we had set $L_{\textrm{max}}$ to $6$ for the entire range
of time, $L_{\textrm{eff}}$ decreased from $6$ to $2$ in recent
years. Since only effective patterns affect the identification of
causal states and transitions among them, they contribute to predict
the future and also can be interpreted as the correlation interval.
Therefore, in the early 1980's one had to look back $6$ ticks for
the prediction of the next tick index, that is, about $6$
consecutive tick indices are correlated. On the other hand in recent
years, one only need to look back $2$ ticks for the prediction,
which means the shorter correlation than before.

The correlation interval is closely related to the time scale for
new information to be delivered to the economic agents and applied
to the market prices \cite{Yang2006,Kaizoji2004}. The decreasing
correlation interval for last twenty years supports that the
information flows faster than before and that the memory length for
the optimal prediction becomes smaller.

\section{Conclusions}

In this paper, we investigated the S\&P500 index from February 1983
to April 2006 by constructing $\epsilon$-machines to infer the
hidden causal structures embedded in the data and by measuring the
statistical and topological complexities from the
$\epsilon$-machines. If in the constructed causal structure the
current causal state is identified, then by following the path from
state to state one can predict the future behavior of a finite
interval. This would be useful in practice, for example, for traders
in financial markets.

We also found that the statistical complexity and the number of
causal states of constructed $\epsilon$-machines have decreased for
last twenty years. Precisely, the length of effective patterns in
histories has become shorter in recent years than in the early
1980's. These results imply that the information flows faster and
hence the memory length needed to predict the future optimally has
become shorter.

% \section{}
% \label{}

% \appendix

% \section{}
% \label{}

% Bibliographic references with the natbib package:
% Parenthetical: \citep{Bai92} produces (Bailyn 1992).
% Textual: \citet{Bai95} produces Bailyn et al. (1995).
% An affix and part of a reference:
%   \citep[e.g.][Ch. 2]{Bar76}
%   produces (e.g. Barnes et al. 1976, Ch. 2).

\newpage

\begin{table}[!ht]
\begin{tabular}{||c|l|l||c|l|l||}
\hline
Causal state& Histories $\overleftarrow{x}$ & Morph &
Causal state& Histories $\overleftarrow{x}$ & Morph \\
name && $\Pr(1 \mid \overleftarrow{x})$ & name && $\Pr(1 \mid
\overleftarrow{x})$\\ \hline
0&000101 100101&0.350333&20&001000&0.39385\\
\cline{4-6}
&100000& &21&010011 110011& 0.434972\\
\cline{1-6}
1&000110 100110&0.507426&22&011011&0.579624\\
\cline{1-6}
2&100001&0.272548&23&010110&0.603976\\
\cline{1-6}
3&001100 101100&0.577508&24&110100&0.559913\\
\cline{1-6}
4&001110 011110 &0.666582&25&111000&0.612022\\
\cline{4-6}
&101110&&26&101010&0.556686\\
\cline{1-6}
5&001111 101111&0.632474&27&011010 111010&0.646215\\
\cline{1-6}
6&111110&0.764303&28&011101 111101&0.641124\\
\cline{1-6}
7&011111&0.712529&29&011100 111100&0.652422\\
\cline{1-6}
8&000010 100010&0.34462&30&110111&0.632246\\
\cline{1-6}
9&001011&0.387701&31&110110&0.670536\\
\cline{1-6}
10&000100&0.362227&32&111111&0.79733\\
\cline{1-6}
11&000011 100011&0.302916&33&010000&0.373832\\
\cline{1-6}
12&001001 101001&0.352661&34&010001 110001&0.345015\\
\cline{1-6}
13&000111 100111&0.48286&35&001101 101101&0.505\\
\cline{1-6}
14&001010&0.492277&36&011000&0.518223\\
\cline{1-6}
15&110000&0.437247&37&010010 110010&0.49167\\
\cline{1-6}
16&100100&0.479832&38&011001 111001&0.483812\\
\cline{1-6}
17&101000&0.480645&39&010100&0.510574\\
\cline{1-6}
18&101011&0.494925&40&010111&0.577414\\
\cline{1-6}
19&010101 110101&0.485992&41&111011&0.636036\\
\hline
\end{tabular}
\caption{The causal states of $\epsilon$-machine constructed from
the February 1983 data. Each causal state consists of its name,
histories, and morph.}
\end{table}

\begin{table}[!ht]
\begin{tabular}{||c|l|l||c|l|l||}
\hline
Causal state& Histories $\overleftarrow{x}$ & Morph &Causal state& Histories $\overleftarrow{x}$& Morph \\
name &  &$\Pr(1 \mid \overleftarrow{x})$ &name &&$\Pr(1 \mid
\overleftarrow{x})$\\
\hline
        0&000011 100011&0.634742&2&000001 100001&0.60027\\
        &000111 100111&&&000101 100101&\\
        &001011 101011&&&001001 101001&\\
        &001111 101111&&&001101 101101&\\
        &010011 110011&&&010001 110001&\\
        &010111 110111&&&010101 110101&\\
        &011011 111011&&&011001 111001&\\
        &011111 111111&&&011101 111101&\\
\hline
        1&000010 100010&0.402178&3&000000 100000&0.36514\\
        &000110 100110&&&000100 100100&\\
        &001010 101010&&&001000 101000&\\
        &001110 101110&&&001100 101100&\\
        &010010 110010&&&010000 110000&\\
        &010110 110110&&&010100 110100&\\
        &011010 111010&&&011000 111000&\\
        &011110 111110&&&011100 111100&\\
\hline
\end{tabular}
\caption{The causal states of $\epsilon$-machine constructed from
the April 2005 data.}
\end{table}

\begin{figure}[!h]
    \centerline{\includegraphics[width=10cm]{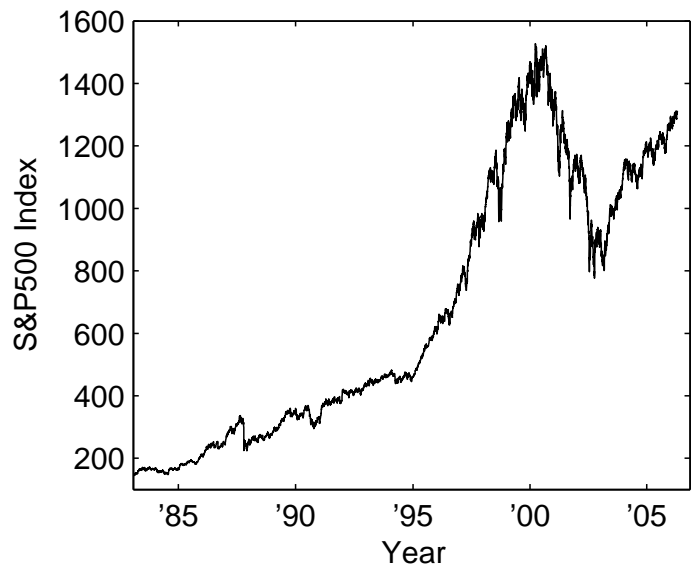}}
\caption{The time series of S\&P500 index from February 1983 to
April 2006.} \label{fig1}
\end{figure}
\begin{figure}[!h]
    \centerline{\includegraphics[width=15cm]{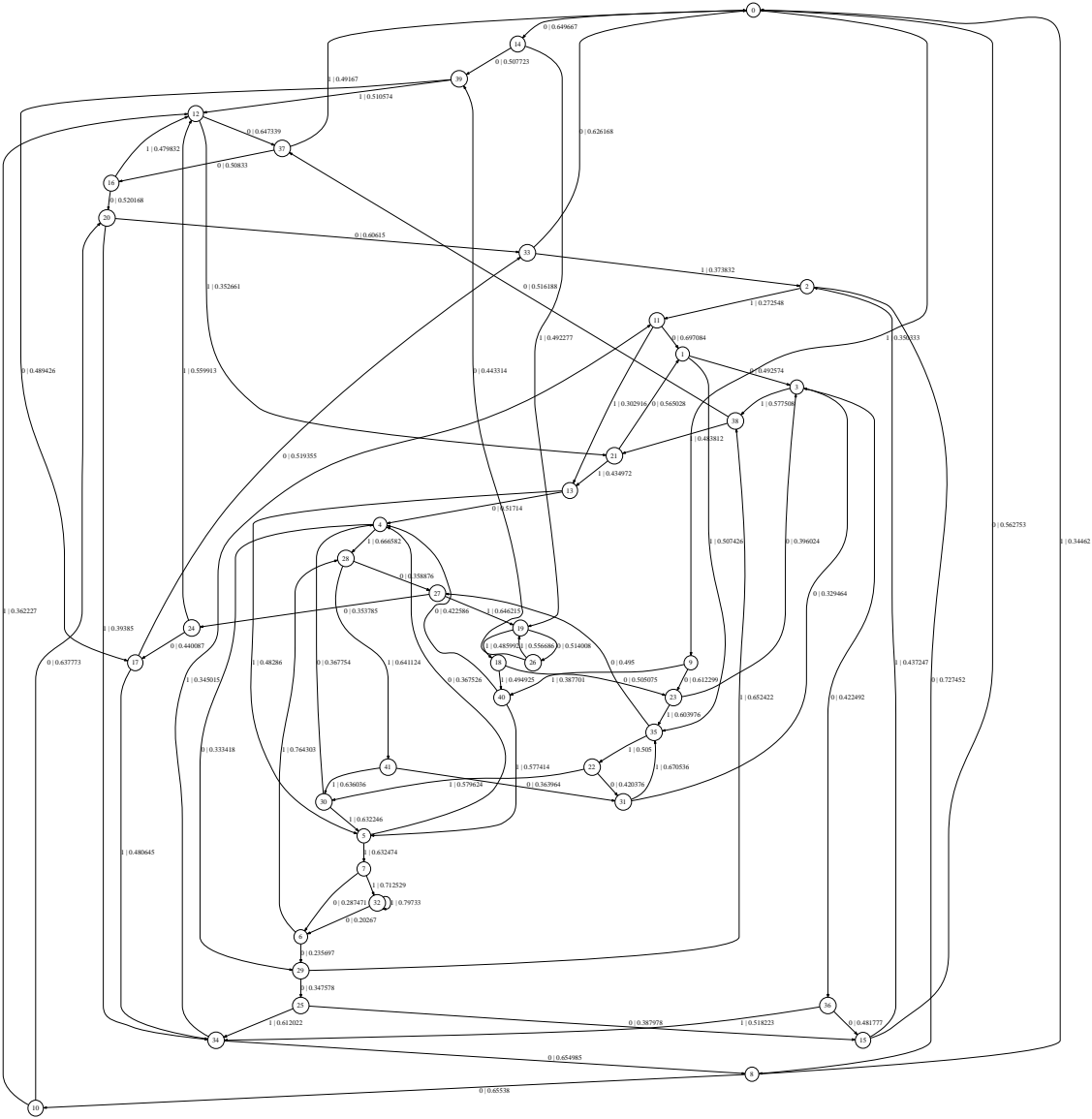}}
    \caption{The $\epsilon$-machine constructed from the February 1983 data of
    S\&P500 index. The figure has been produced with the Graphviz software (http://www.graphviz.org/).} \label{fig2}
\end{figure}
\begin{figure}[!h]
    \centerline{\includegraphics[width=10cm]{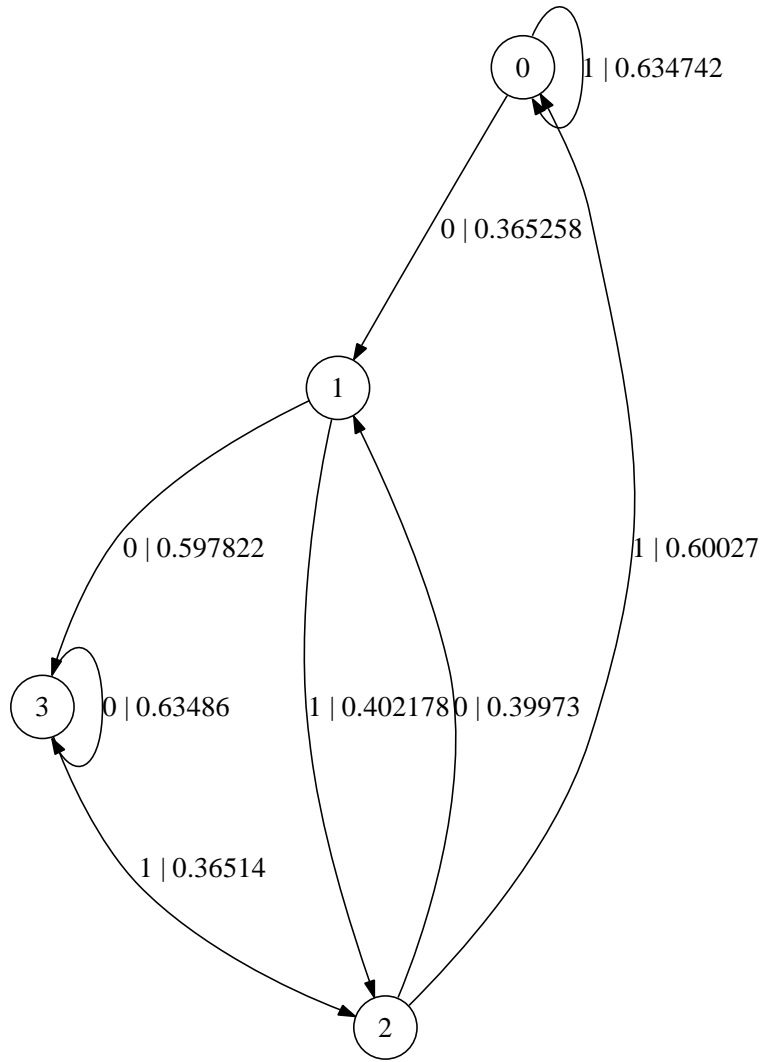}}
    \caption{The $\epsilon$-machine constructed from the April 2005 data of S\&P500 index. The figure has been produced with the Graphviz software.}
\label{fig3}
\end{figure}
\begin{figure}[!h]
    \centerline{\includegraphics[width=10cm]{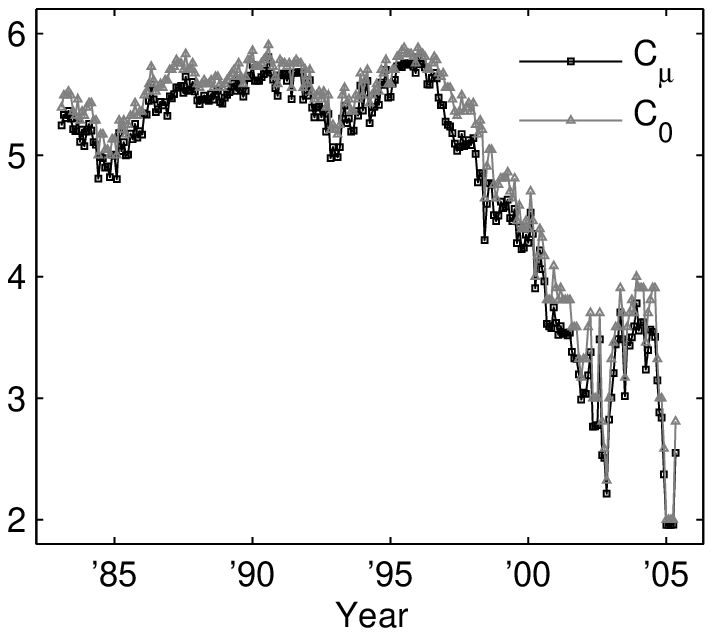}}
    \caption{The decreasing behaviors of the statistical complexity $C_\mu$ and the topological complexity $C_0$
of S\&P500 index from February 1983 to April 2006. Notice that $C_0$
is upper bound of $C_{\mu}$.} \label{fig4}
\end{figure}


\begin{thebibliography}{00}

% \bibitem[Names(Year)]{label} or \bibitem[Names(Year)Long names]{label}.
% (\harvarditem{Name}{Year}{label} is also supported.)

\bibitem{Arthur1997} W. B. Arthur, S. N. Durlauf, D. A. Lane (Eds.),
The Economy as an Evolving Complex System II, Perseus Books, 1997.

\bibitem{Mantegna2000} R. N. Mantegna, H. E. Stanley, An Introduction
to Econophysics: Correlations and Complexity in Finance, Cambridge
University Press, 2000.

\bibitem{Bouchaud2000} J.-P. Bouchaud, M. Potters, Theory of
Financial Risks: From Statistical Physics to Risk Management,
Cambridge University Press, 2000.

\bibitem{Mandelbrot2001} B. B. Mandelbrot, Quant. Finance 1 (2001)
124-130.

\bibitem{Kaizoji2006} T. Kaizoji, On Stock-Price Fluctuations in the
Periods of Booms and Stagnations, in: A. Chatterjee, B. K.
Chakrabarti (Eds.), Econophysics of Stock and Other Markets,
Springer-Verlag Italia, Milan, 2006.

\bibitem{Kullmann2000} L. Kullmann, J. Kert\'esz, R. N. Mantegna,
Physica A 287 (2000) 412-419.

\bibitem{Challet1997} D. Challet, Y.-C. Zhang, Physica A 246 (1997)
407-418.

\bibitem{Scalas2000} E. Scalas, R. Gorenflo, F. Mainardi, Physica A
284 (2000) 376-384.

\bibitem{Giada2002} L. Giada, M. Marsili, Physica A 315 (2002)
650-664.

\bibitem{Peters1991} E. E. Peters, Chaos and order in the capital
markets, Wiely, 1991.

\bibitem{Peng1994} C.-K. Peng, S. V. Buldyrev, S. Havlin, M. Simons,
H. E. Stanley, A. L. Goldberger, Phys. Rev. E 49 (1994) 1685-1689.

\bibitem{Liu1999} Y. Liu, P. Gopikrishnan, P. Cizeau, M. Meyer, C.-K.
Peng, H. E. Stanley, Phys. Rev. E 60 (1999) 1390-1400.

\bibitem{Feldman1998} D. Feldman, A Brief Introduction to Information
Theory, Excess Entropy and Computational Mechanics,
http://hornacek.coa.edu/dave/Tutorial/, April 1998.

\bibitem{Shalizi2001} C. R. Shalizi, J. P. Crutchfield, J. Stat.
Phys. 104 (2001) 817-879.

\bibitem{Shannon1948} C. E. Shannon, Bell System Tech. J. 27 (1948)
379-423; 623-656.

\bibitem{Kolmogorov1965} A. N. Kolmogorov, Problems of Information
Transmission 1 (1965) 1-7.

\bibitem{Chaitin1966} G. J. Chaitin, Journal of the ACM 13 (1966)
547-569.

\bibitem{Hanson1997} J. E. Hanson, J. P. Crutchfield, Physica D 103
(1997) 169-189.

\bibitem{Shalizi2004} C. R. Shalizi, K. L. Shalizi, R. Haslinger,
Phys. Rev. Lett. 93 (2004) 118701.

\bibitem{Crutchfield1997} J. P. Crutchfield, D. P. Feldman, Phys.
Rev. E 55 (1997) 1239-1242.

\bibitem{Clarke2003} R. W. Clarke, M. P. Freeman, N. W. Watkins,
Phys. Rev. E 67 (2003) 016203.

\bibitem{Palmer2000} A. J. Palmer, C. W. Fairall, W. A. Brewer, IEEE
Transactions on Geoscience and Remote Sensing 38 (2000) 2056-2063.

\bibitem{Crutchfield1989} J. P. Crutchfield, K. Young, Phys. Rev.
Lett. 63 (1989) 105-108.

\bibitem{Shalizi2004b} C. R. Shalizi, K. L. Shalizi, Blind
Construction of Optimal Nonlinear Recursive Predictors for Discrete
Sequences, in: M. Chickering, J. Halpern (Eds.), Proceedings of the
20th Annual Conference on Uncertainty in Artificial Intelligence
(UAI-04), AUAI Press, Virginia, 2004, pp. 504-511.


\bibitem{Feldman1998b} D. P. Feldman, J. P. Crutchfield, Phys. Lett.
A 238 (1998) 244-252.

\bibitem{CSSR} $\epsilon$-machines are reconstruced by the causal
state splitting reconstruction (CSSR) algorithm developed by C. R.
Shalizi and K. Klinkner. For more information visit their homepage:
http://www.cscs.umich.edu/~crshalizi/CSSR/.

\bibitem{Yang2006} J.-S. Yang, S. Chae, W.-S. Jung, H.-T. Moon,
Physica A 363 (2006) 377-382.

\bibitem{Kaizoji2004} T. Kaizoji, Physica A 343 (2004) 662-668.

\end{thebibliography}
\end{document}